\documentclass[a4paper,11pt]{article}
\pdfoutput=1 
\usepackage{jinstpub} 
\usepackage{float}
\usepackage{caption}
\usepackage{subcaption}
\usepackage{amsmath}
\usepackage{siunitx}
\usepackage[normalem]{ulem}
\usepackage{color}
\usepackage[left]{lineno}

\title{Characterisation of Silicon Photomultipliers for Liquid Xenon Detectors}

 \author{Laura~Baudis,}
 \author{Michelle~Galloway,}
 \author{Alexander~Kish,}
 \author{Chris~Marentini,}
 \author{and Julien~Wulf\footnote{Corresponding author.} }
 \affiliation{ Physik-Institut, Universit\"at Z\"urich, \\
Winterthurerstr. 190, CH-8057, Switzerland}

\emailAdd{jwulf@physik.uzh.ch}

\abstract{Silicon Photomultipliers (SiPMs) are considered as a solid-state sensor alternative to photomultiplier tubes in experiments using liquid xenon (LXe) as a radiation detection medium. The main  requirements are single-photon detection of the vacuum ultraviolet scintillation light from LXe at 178\,nm with high  resolution and detection efficiency and low noise rates. Further requirements for dark matter and double beta decay searches are ultra-low radioactivity levels of all the components including the substrates and cold electronics. Here we describe our characterisation of Hamamatsu 6$\times$6\,mm$^2$ SiPMs in the temperature range 110-300\,K in nitrogen gas, as well as long-term measurements in cold nitrogen gas at 172\,K and liquid xenon at 185\,K. After we introduce the experimental setups, the data acquisition schemes and analysis methods, we show the single-photon response, the gain versus bias voltage, as well as the dark and correlated noise rates. We demonstrate the long-term stability at cryogenic temperatures, and conclude that SiPM arrays are promising candidates for photosensor arrays in liquid xenon detectors. Furthermore, we study the radioactivity of the raw SiPM materials with gamma spectrometry and inductively coupled plasma mass spectrometry and conclude that SiPMs are suitable for use in low-background experiments.}

\begin{document}

\maketitle
\flushbottom

\section{Introduction}
\label{sec:introduction}

Rare-event searches using xenon as a radiation detection medium mostly employ  photomultiplier tubes (PMTs) to detect the vacuum ultraviolet (VUV) scintillation photons  caused by particles interacting in the xenon. Examples are the XMASS~\cite{Abe:2013tc}, XENON1T~\cite{Aprile:2017aty}, LUX~\cite{Akerib:2012ys} and PandaX~\cite{Cao:2014jsa} dark matter experiments, while the double-beta decay experiments EXO-200~\cite{Albert:2017hjq} and NEXT~\cite{Ferrario:2017zqp} use large-area avalanche photodiodes (LAAPDs) and silicon photomultipliers (SiPMs) and PMTs, respectively.

Silicon photomultipliers are indeed promising candidates to replace PMTs in future detectors, and are considered, e.g., in nEXO~\cite{Albert:2017hjq} and DARWIN~\cite{Aalbers:2016jon}, as well as in DarkSide-20k~\cite{Aalseth:2017fik}, which will use argon as detection medium. These are arrays of reverse-biased avalanche photodiodes (APDs) operated in Geiger mode and connected in parallel. Each cell has the same geometrical size and is an independently operating unit, which starts an avalanche when a photon is absorbed. SiPMs have several advantages: compact geometry, low operating voltage, simplicity of readout and excellent single photoelectron resolution. Additional expected benefits include low radioactivity levels, low costs, and scalable mass production. 
To successfully employ SiPMs in rare-event searches, low dark count rates are required to minimise the rate of accidental coincidences at low energy thresholds, and a gain in the order of $10^6$ is needed for a high single photoelectron (SPE) detection efficiency.

Here we evaluate a single 6$\times$6\,mm$^2$ segment of the S13371 (VUV4) 12$\times$12\,mm$^2$ VUV sensitive SiPM from Hamamatsu. This device is optimised for operation in experiments using xenon as target, and was first developed for the MEG-II~\cite{Ogawa:2017xof} experiment at Paul Scherrer Institute (PSI).  In section \ref{sec:characterisation} we describe our experimental setups and the data acquisition system, as well as the analysis method. We then show our results for the gain, dark count rate and crosstalk rate as a function of temperature. We conclude the section with results from long-term measurements of the gain and dark count rate at a temperature of 172\,K. In section \ref{sec:liquid_xenon_operation} we present the results from the SiPM operation in liquid xenon, including measurements with $^{83\mathrm{m}}$Kr and $^{241}$Am calibration sources. 
In section \ref{sec:Radiopurity} we detail a study of the radioactivity of the SiPM raw materials from Hamamatsu, using gamma spectrometry and inductively coupled plasma mass spectrometry.
We close with a summary of our main findings and conclusions  in section~\ref{sec:conclusions}. 


\section{Temperature Dependent SiPM Characteristics}
\label{sec:characterisation}

In this section we first describe the experimental setup which was used to characterise the photosensor. This is followed by a short description of the SiPM, its readout scheme and the analysis procedure of raw traces. We then show the results of the SiPM characteristics as a function of bias voltage $\left( V_{\mathrm{bias}} \right)$ at different temperatures in gaseous nitrogen. We conclude with results of a long-term measurement at 172\,K in gaseous nitrogen.  


\subsection{Experimental Setup and Data Acquisition System}
\label{subsection:setup}

\begin{figure}[b!]
\centering
\includegraphics[height=6.5cm]{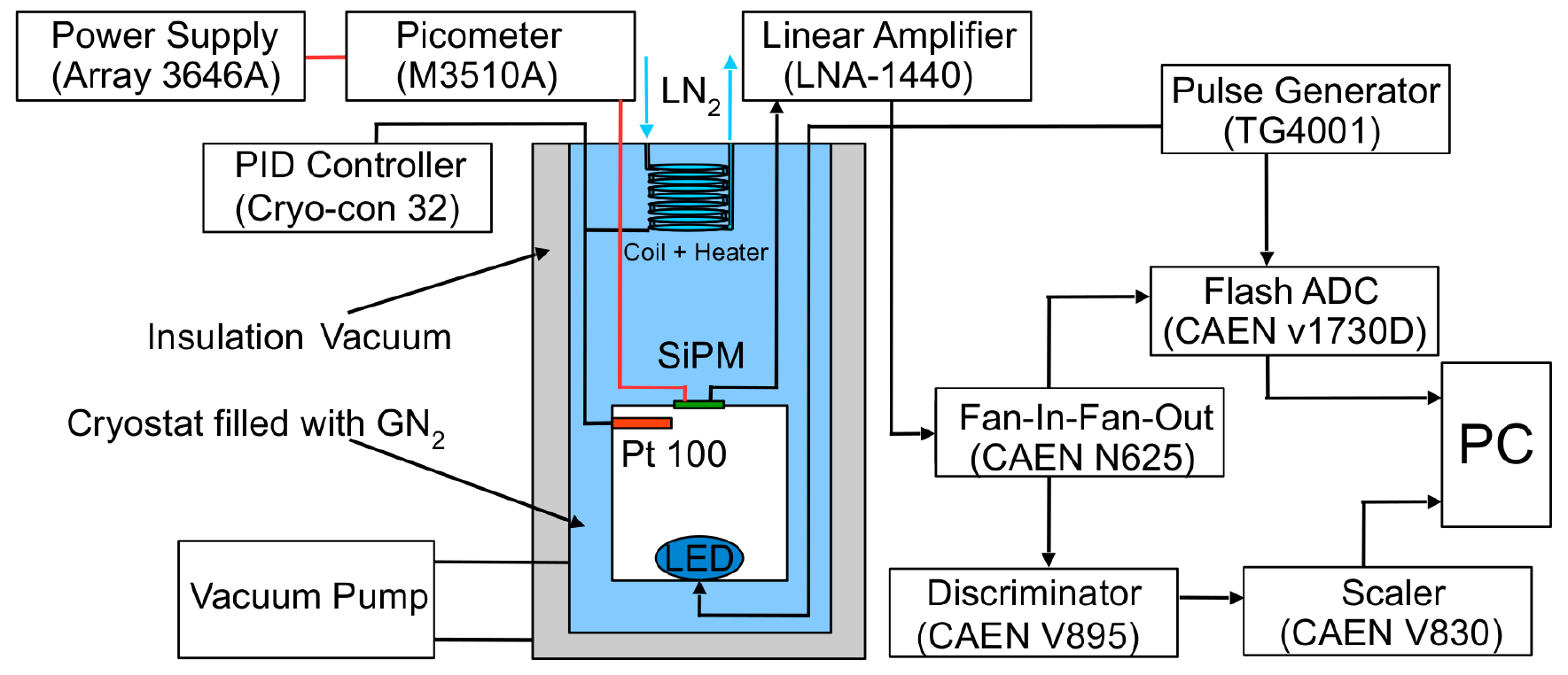}
\caption{A sketch of the experimental apparatus and the data acquisition system. The SiPM is positioned near the centre of the cryostat on a PTFE holder, and the cryostat is filled with gaseous nitrogen. Liquid nitrogen is flushed through the copper coil and the temperature is controlled via a heating foil wrapped around the coil and regulated by a PID controller. } 
\label{fig:setup}
\end{figure}

A sketch of the experimental apparatus employed in all the temperature dependent measurements is shown in Figure \ref{fig:setup}. The setup consists of an inner and outer chamber. Within the inner chamber, either a PMT or several SiPMs can be supported by a polytetrafluoroethylene  (PTFE) holder. The temperature is measured by two Pt-100 resistant thermometers placed at 2\,mm distance to the SiPM inside the holder. The inner volume is first evacuated using a turbo molecular pump to a pressure below $10^{-4}$\,mbar and filled with gaseous nitrogen with purity level of 99.9999\% (6.0) to avoid condensation of water or other gases. The nitrogen pressure within the chamber is maintained at 
$\sim$\,1.8\,bar. Liquid nitrogen is flushed though a copper cooling coil to reduce the temperature of the chamber, and the cooling power is maintained by setting a constant flow rate at the outlet of the coil. A heating foil wrapped around the cooling coil and regulated by a PID controller (Cryo-con model 32) allows us to control the temperature. Prior to every measurement at each temperature setting, we waited for 30\,mins until the temperature was stable within $\pm$0.02\,K.

The SiPM is positioned near the centre of the cryostat on the PTFE holder, facing a blue LED ($\lambda\sim$\,470\,nm). A pulse generator (TTi TG5012A) is used to bias the LED with a square pulse and to simultaneously  trigger  the  data  acquisition system. An external power supply (Array 3646A) provides the bias voltage for the SiPM  and a pico-ampere-meter (M3510A) is used to measure the corresponding current. The signal from the SiPM is amplified by a 100-fold commercial low noise amplifier (RF Bay LNA-1440) and duplicated with a Quad Linear Fan-in Fan-out (CAEN N625). The first signal copy is digitised by a flash ADC (CAEN V1730D) with a sampling frequency of 500 MHz, a bandwidth of 250 MHz and a resolution of 14 bits, for subsequent analysis. The second copy is fed into a discriminator (CAEN V895) with a programmable threshold, and the resulting trigger output is propagated to a scaler (CAEN V830). This configuration allows us to count the number of signals above the programmable threshold in a certain time window.

\subsection{SiPM Description and Signal Trace Analysis}
\label{subsection:processors}

\begin{figure}
\centering
\includegraphics[height=5.3cm]{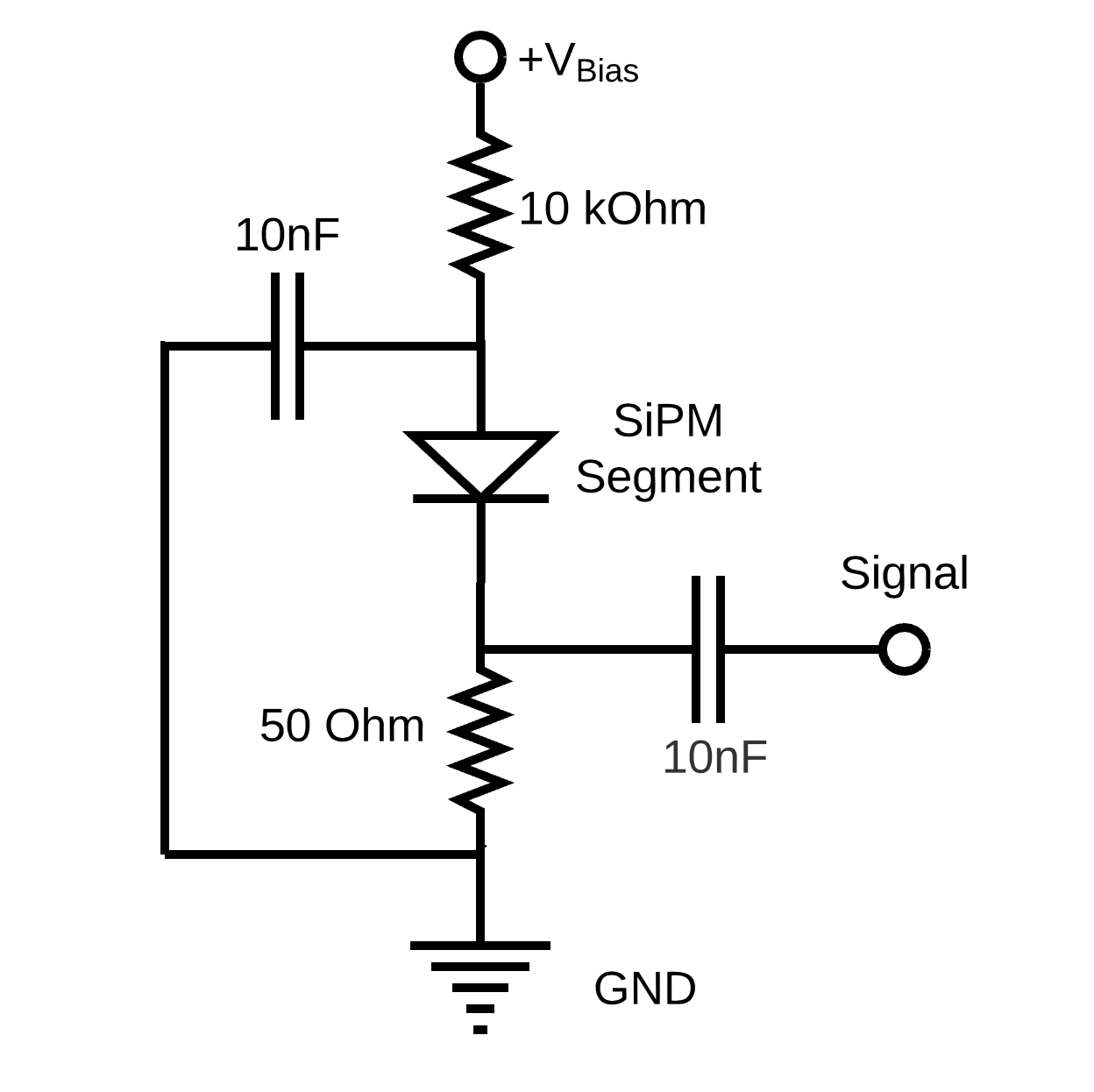}
\hspace{0.2cm}
\includegraphics[height=5.7cm]{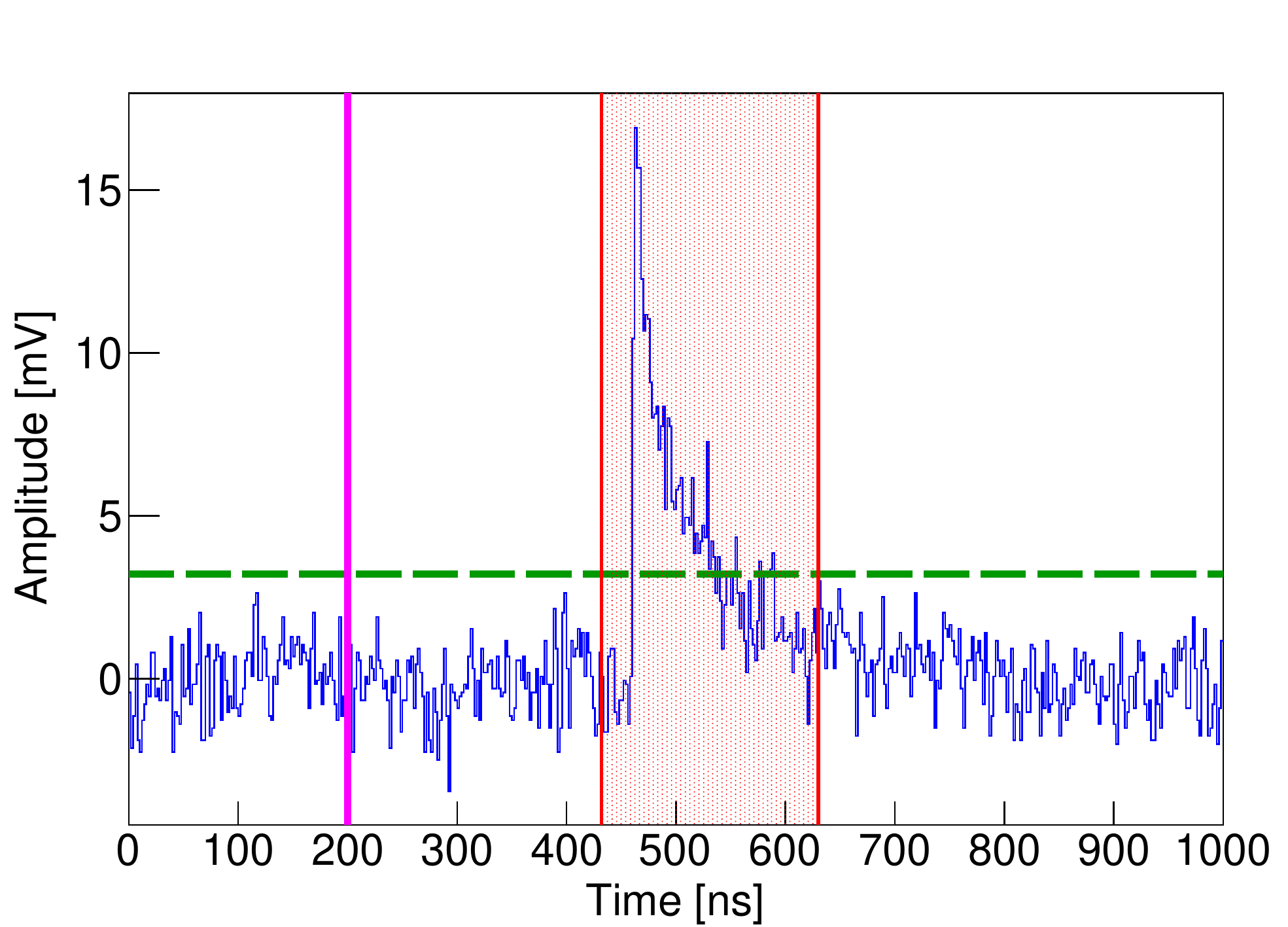}
\caption{(Left): The readout schematic of the SiPM. The SiPM is biased with a positive voltage and the signal pulse is read out with a capacitor. (Right): The digitised voltage pulse from the SiPM. The data processor integrates the pulse shape (red shaded area) above three times the calculated rms of the baseline (green dashed line) in the red shaded region. The baseline is calculated within the first 200\,ns of the waveform (magenta line).}
\label{fig:SiPM_Trace}
\end{figure}

Most commercially available SiPMs are sensitive to light in the visible range of the electromagnetic spectrum, which makes them unsuitable for noble gas detectors, unless a wavelength shifter to convert the VUV photons into the visible regime is used. The SiPM from Hamamatsu with an active area of $6\times6$\,mm$^2$ and a single cell size of 50\,$\mu$m $\times$ 50\,$\mu$m  is sensitive to xenon scintillation light at 178\,nm with a photo-detection efficiency around 24\%~\cite{Hamamatsu:2018}. 

The SiPM is connected on the cathode side to the bias voltage in series with a 10\,k$\Omega$ resistor to limit the current, in order not to damage the device. The anode side is grounded with a  50\,$\Omega$ resistor and capacitively coupled to an external voltage amplifier. The readout schematic is shown in Figure \ref{fig:SiPM_Trace}, left. More information on a possible readout of the SiPM can be found in \cite{sandro2016}. The generated voltage pulses from the SiPM are digitised and stored for subsequent processing.  A data processor scans the stored waveforms for excursions which are above three times the rms of the baseline, calculated within the first 200\,ns. If this threshold is exceeded by a pulse from the SiPM, the pulse shape is integrated. The calculated integral area corresponds to the signal charge produced by the discharge of the individual cells in the SiPM. An example of the output of the data processor is displayed in Figure \ref{fig:SiPM_Trace}, right. The area of the pulse, its height, width, position, decay time and other parameters, such as the calculated threshold, the number of detected pulses, the baseline and the baseline rms are stored for further analysis.

\subsection{Single Photon Response, Breakdown Voltage and I-V Characteristics}
\label{subsection:spe}

\begin{figure}[b!]
\centering
\includegraphics[height=5.3cm]{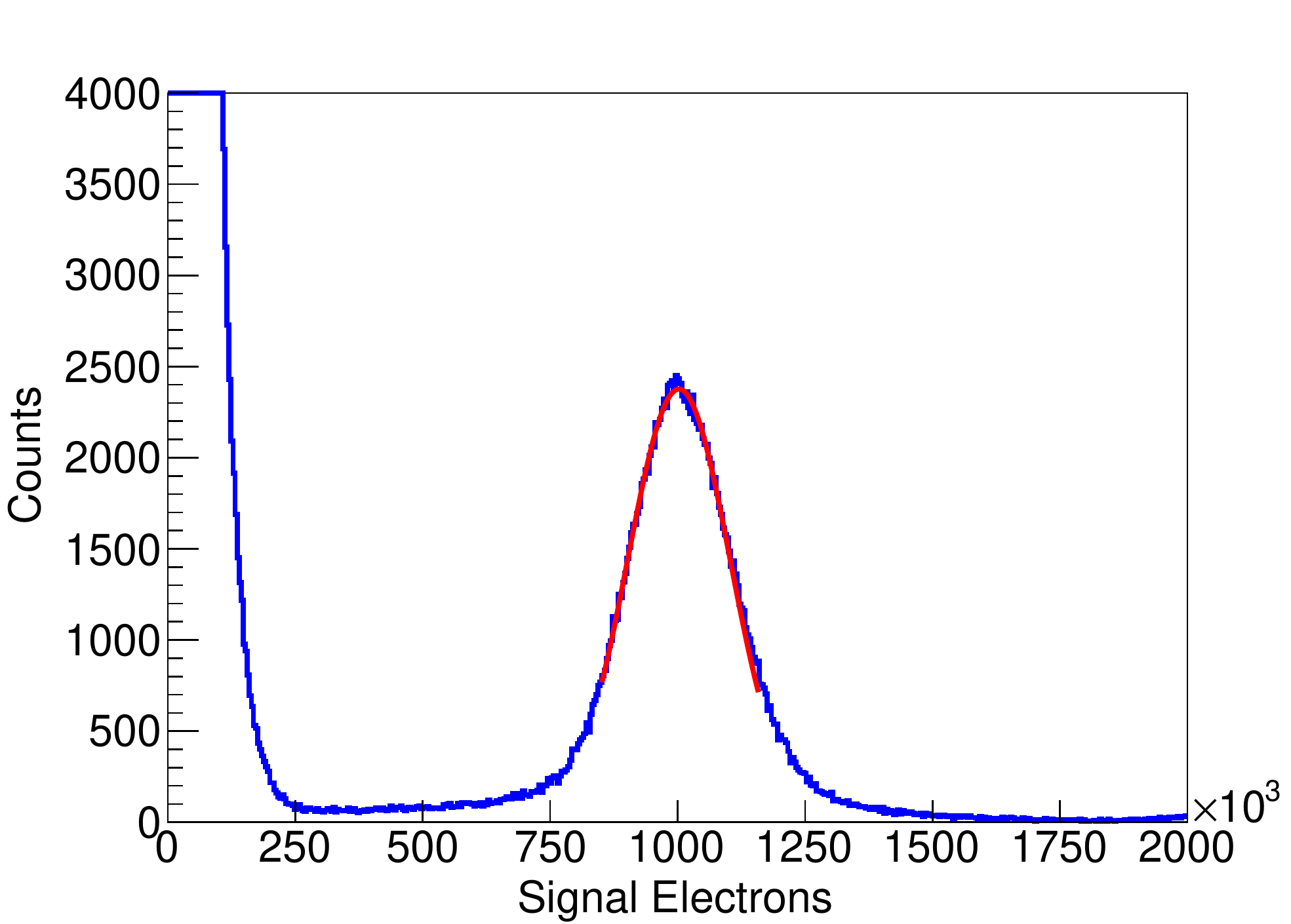}
\includegraphics[height=5.3cm]{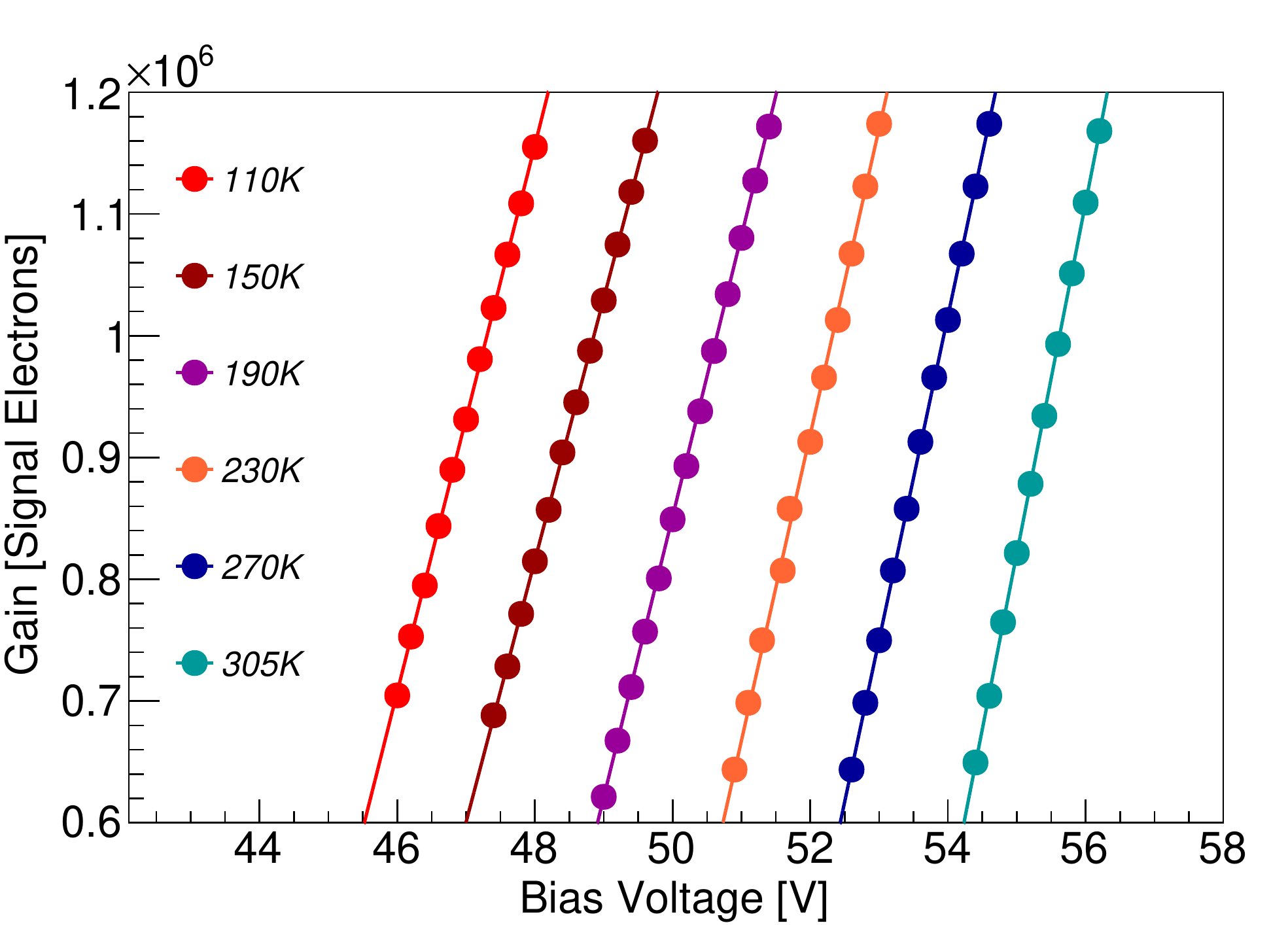}
\caption{(Left): Example of a single photoelectron spectrum (blue curve) at a bias voltage of 54.0\,V and an operation temperature of 270\,K.  The photoelectron peak is fitted with a Gaussian function (red curve) to obtain the gain. (Right): The gain as a function of bias voltage  at different temperatures in gaseous nitrogen. The gain data points at each temperature setting are fitted with a linear function in order to obtain the breakdown voltage as the intersection of the fit with the x-axis.}
\label{fig:SPE_IV}
\end{figure}

\begin{figure}[btp]
\centering
\includegraphics[height=5.3cm]{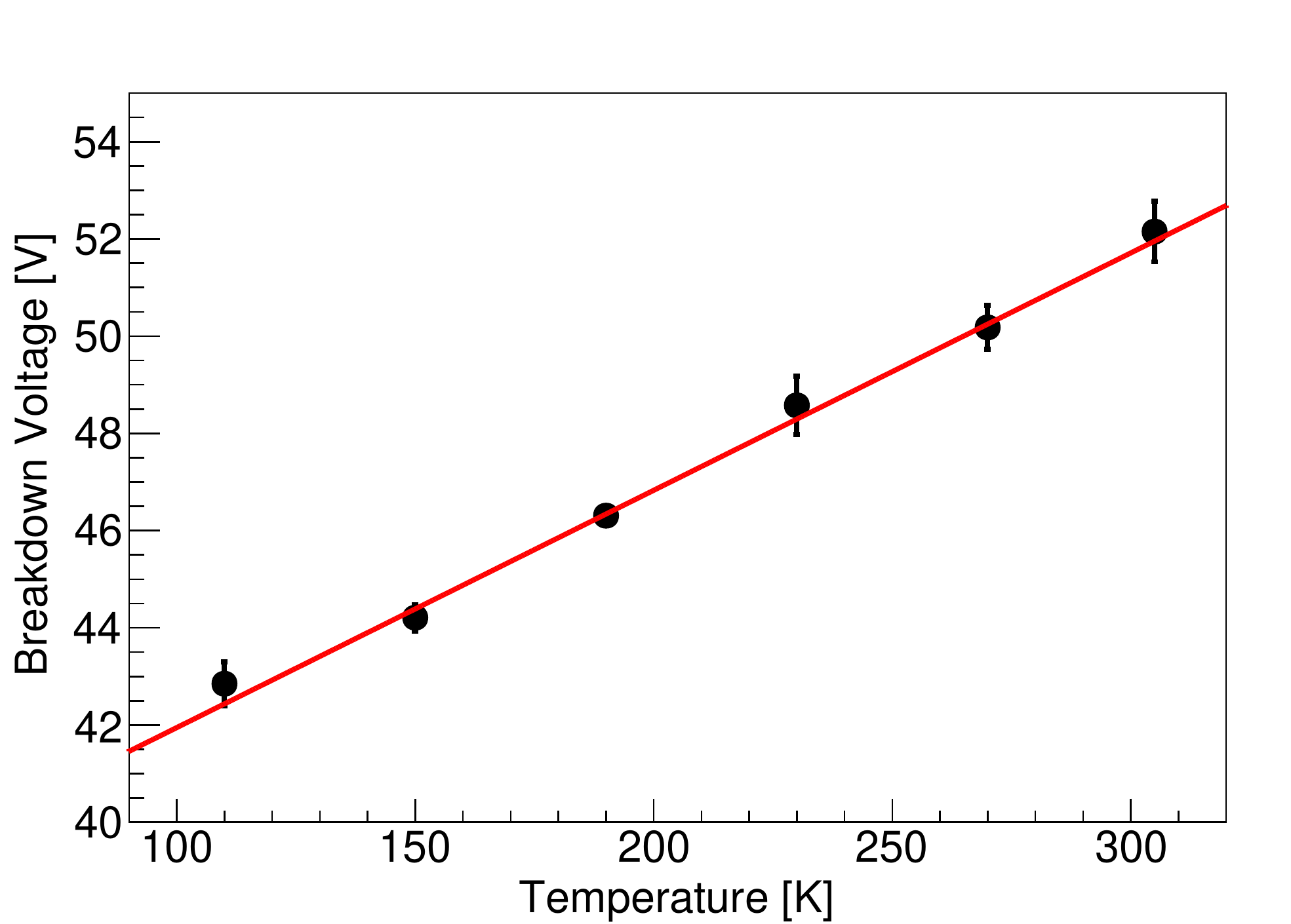}
\includegraphics[height=5.3cm]{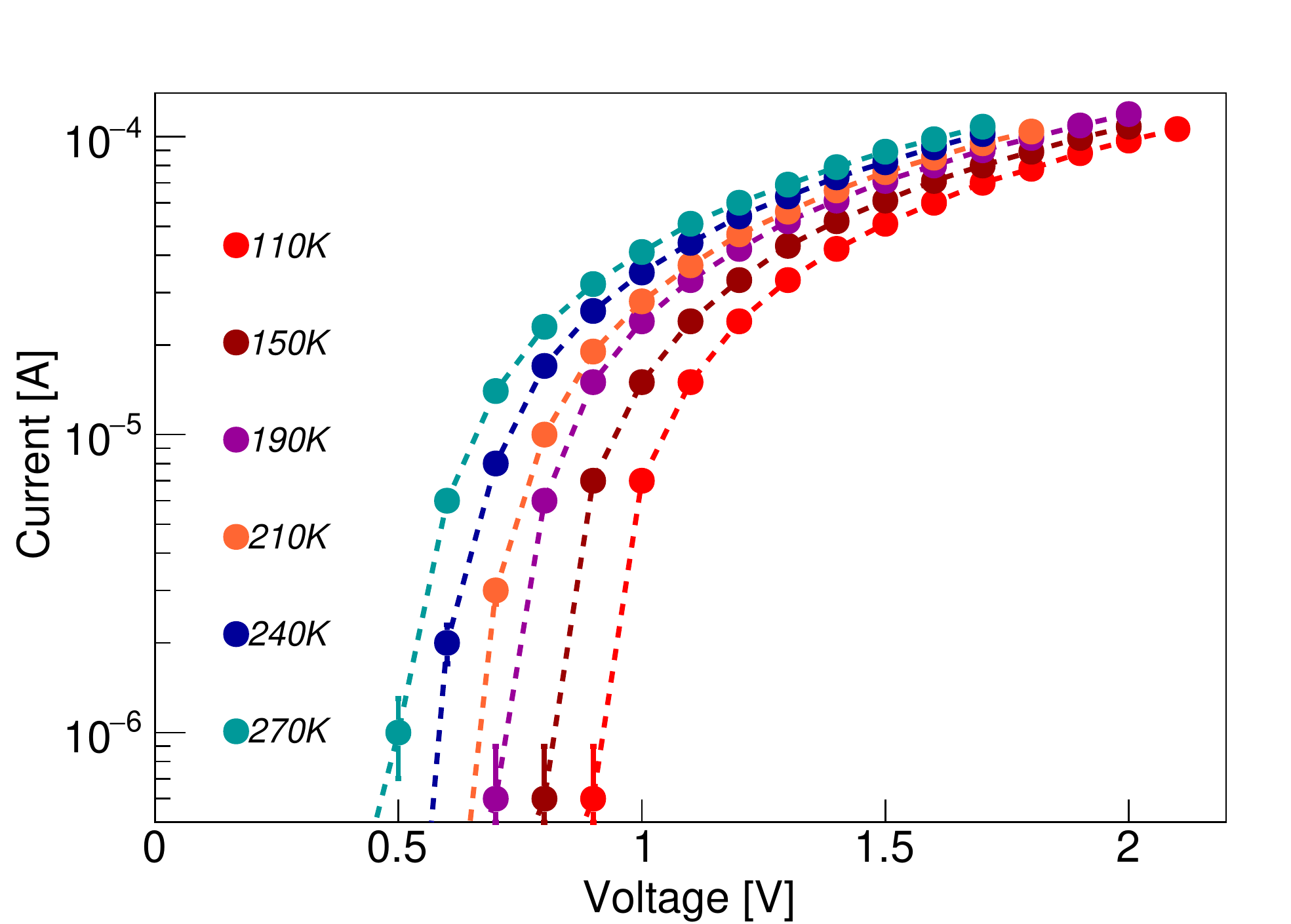}
\caption{(Left): The breakdown voltage as a function of temperature (black data points) along with a linear fit (red line). The breakdown voltage decreases with decreasing temperature  with the slope $ \left( 48.6 \pm 0.4 \right)$\,mV per Kelvin. (Right): The forward biased I-V curve at different temperatures. The SiPM resistance is obtained by fitting the curves with a linear function at high voltages, and  increases slightly with decreasing temperature.}
\label{fig:gain_over_t}
\end{figure}

To measure the gain, resolution and breakdown voltage, a single photoelectron spectrum is first acquired by illuminating the SiPM with light from a blue LED. It is triggered by the external pulse generator with a frequency of 1\,kHz and provides an illumination rate of $\sim$\,1\,photon for every trigger. For every bias voltage,  500\,000 events were acquired. Each digitised event captured a window of 1\,$\mu$s and the signal traces were stored on a computer. A set of data quality cuts have been applied to the acquired waveforms, in order to reject events which can not be analysed due to misidentified pulses, shifted baselines, correlated noise, or other conditions. The gain was obtained from the single photoelectron (SPE) spectrum by fitting the photoelectron peak with a Gaussian function, as shown in Figure \ref{fig:SPE_IV}, left. 
The mean $\left(\mu_{\mathrm{SPE}}\right)$ of the Gaussian  corresponds to the gain, and the ratio between the sigma $\left(\sigma_{\mathrm{SPE}}\right)$ and the mean $\mu_{\mathrm{SPE}}$ to the resolution.  In Figure \ref{fig:SPE_IV}, right, we show the obtained gains between 110\,K$-$305\,K as a function of bias voltage. The collected data points were fitted with a linear function to obtain the breakdown voltage at each temperature setting, which lies at the intersection of the linear functions with the x-axis at zero gain. The breakdown voltage decreases linearly towards lower temperatures with $\left(48.6\pm0.4\right)$\,mV per Kelvin, as shown in Figure~\ref{fig:gain_over_t}, left, by the slope of the linear fit to the data. The SPE resolution improves with over-voltage ($\Delta V = V_{\mathrm{bias} } - V_{\mathrm{break} }$) for all temperatures to the lowest measured value of  $\sim$9\%. This value is superior to cryogenic PMTs, which show an SPE resolution around 25$-$30\% \cite{Barrow:2016doe}. 

The produced charge during the avalanche process is equivalent to the product of the capacitance and the applied over-voltage ($Q=C\times \Delta V$), the SiPM cell capacitance is obtained from the slope of the fitted linear function. We measured a SiPM cell capacitance of $(369.5\pm0.5)$\,fF at 190\,K. We forward biased the SiPM at different temperatures to determine the resistance of the SiPM from the linear dependency between voltage and current at high currents, as shown in Figure \ref{fig:gain_over_t}, right. The  resistance varies from $\sim$\,$10.7\,\mathrm{k}\Omega$ at 270\,K up to $\sim$\,$11\,\mathrm{k}\Omega$ at 110\,K. The slight increase of the SiPM resistance has therefore no effect on the pulse shape of the SiPM.

\subsection{Dark Count and Crosstalk Rates}
\label{subsection:dcr}

The SiPM exhibits several noise characteristics, which can be grouped into uncorrelated noise, such as thermally generated dark counts, and pixel discharging correlated noise, such as crosstalk. The dark count rate (DCR) was measured by counting thermally generated single pulses at different bias voltages as a function of temperature above a 0.5\,PE threshold ($\mathrm{DCR}_{0.5\,\mathrm{PE}}$). These measurements were performed at temperatures above 230\,K, for lower fluctuations in the dark count rate.

\begin{figure}[h!]
\centering
\includegraphics[height=5.3cm]{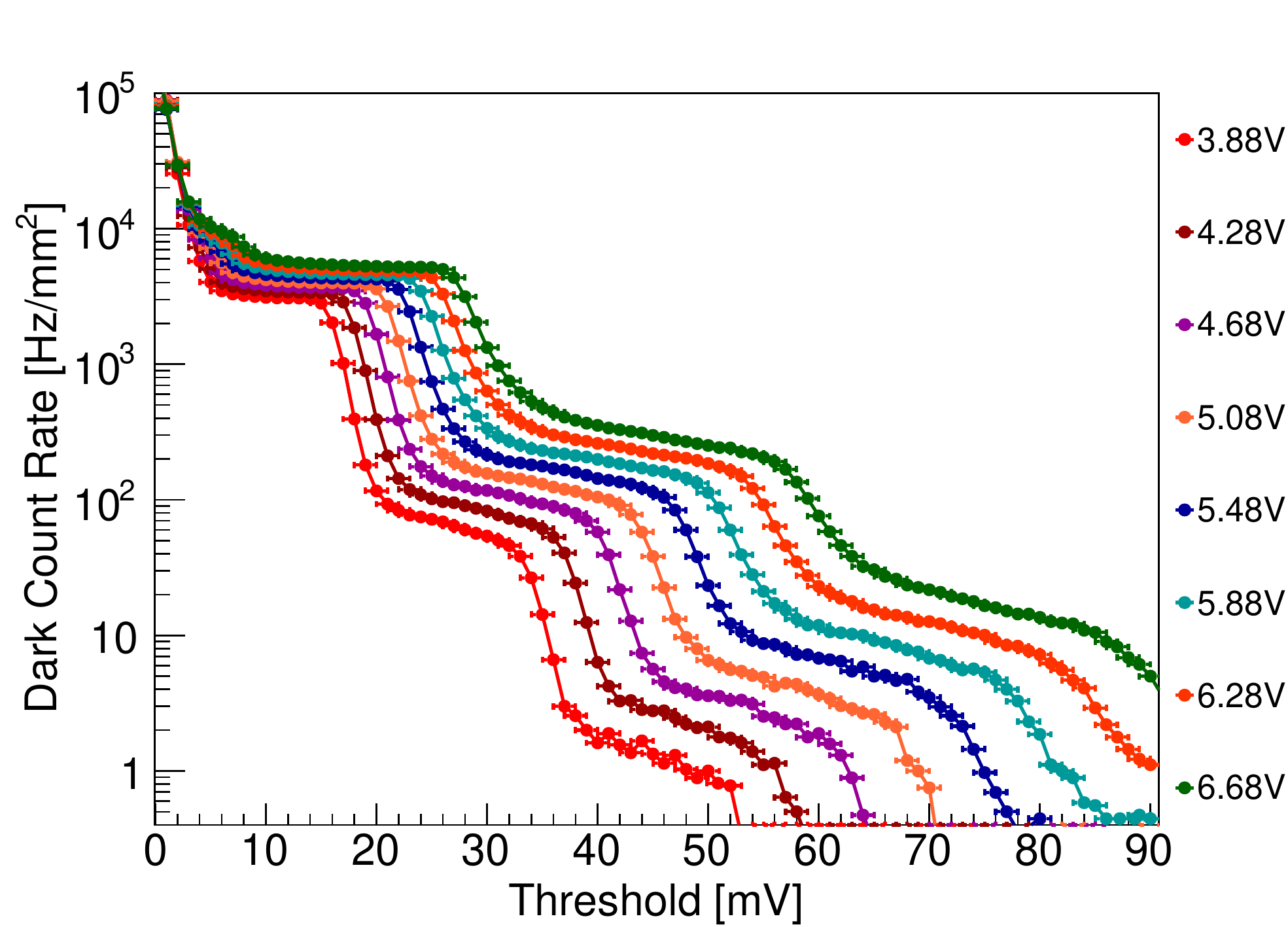}
\includegraphics[height=5.3cm]{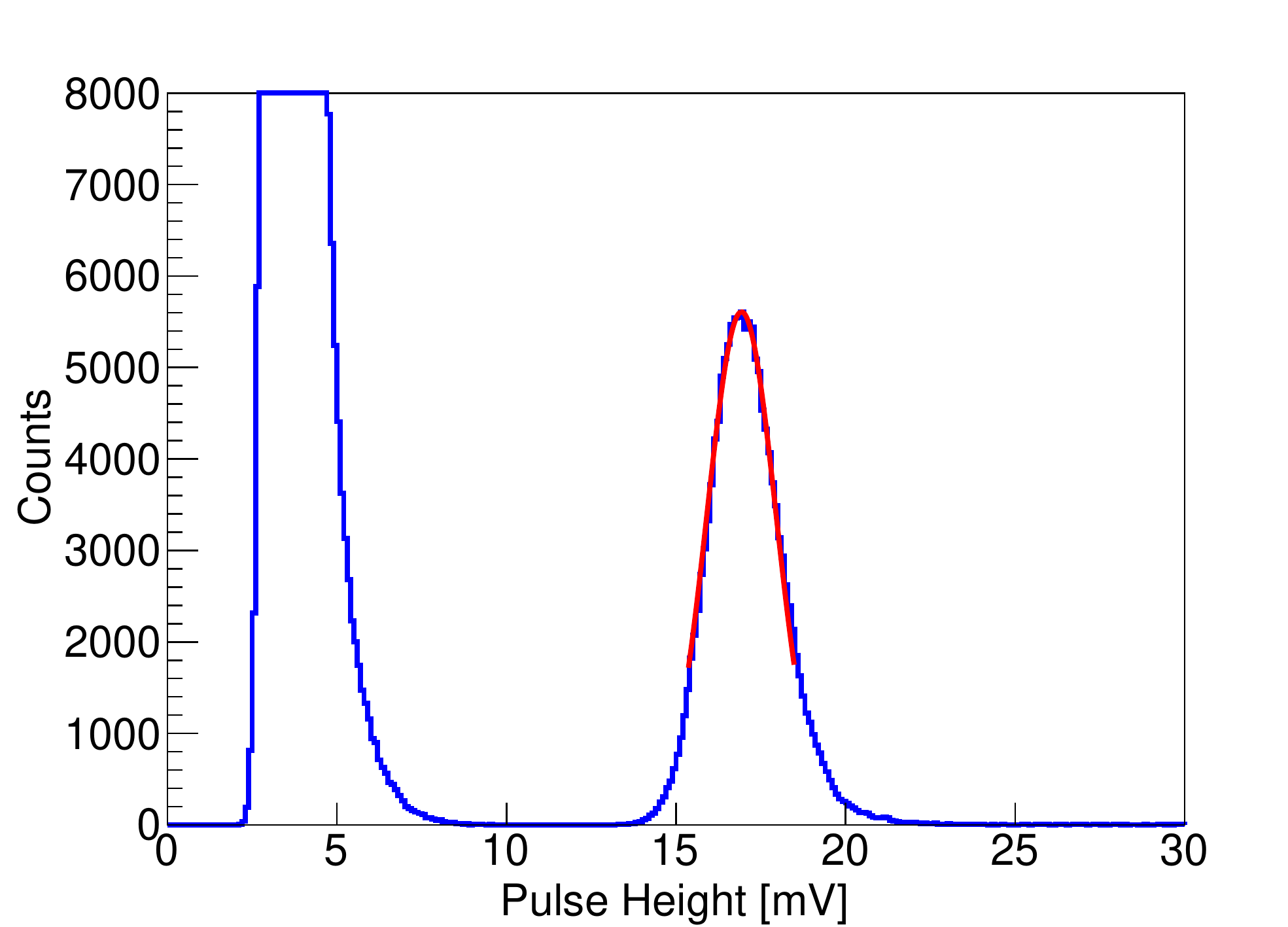}
\caption{(Left): Dark count rate as a function of signal threshold for different over-voltages at 260\,K. The behaviour is characterised by a series of plateaus, which correspond to 0.5\,PE, 1.5\,PE and 2.5\,PE pulse heights. (Right): Example of a pulse height histogram (blue curve), where the single photoelectron peak is fitted with a Gaussian function (red curve). The mean value is used to calculate the 0.5\,PE and 1.5\,PE thresholds.  } 
\label{fig:DCR_Sweeping}
\end{figure}

\begin{figure}[t!]
\centering
\includegraphics[height=5.3cm]{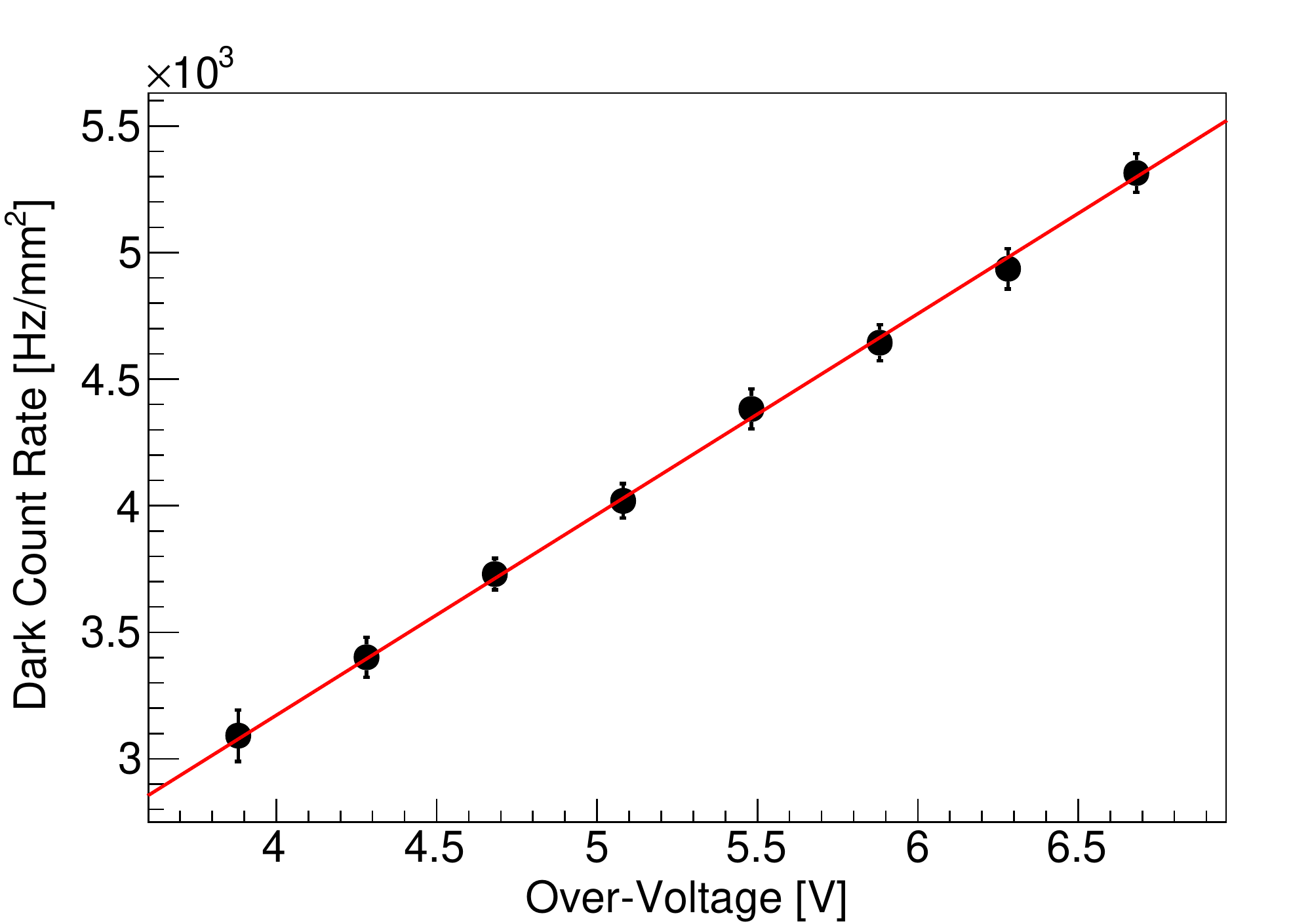}
\includegraphics[height=5.3cm]{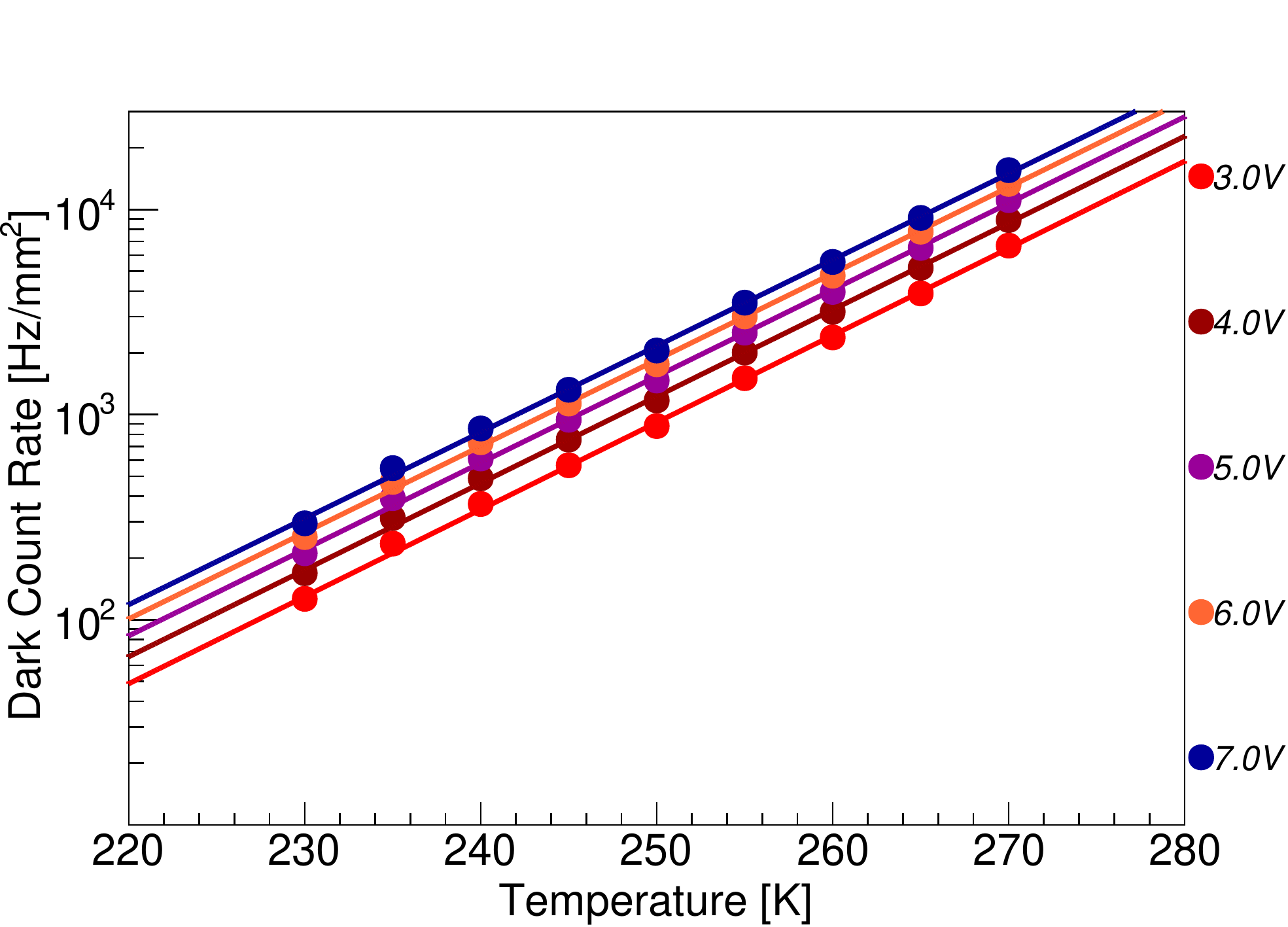}
\caption{ (Left): Dark count rate as a function of over-voltage at 260\,K. The points are fitted with a linear function (red line). (Right): The dark count rate as a function of temperature for different over-voltages from 3 to 7\,V. The rates follow an exponential decrease over two orders of magnitude and are fitted with an exponential function, see text.}
\label{fig:DC_Results}
\end{figure}

At each temperature setting, the SiPM dark count rate was measured by sweeping through different signal thresholds. Figure \ref{fig:DCR_Sweeping}, left, shows the DCR for different \mbox{over-voltages} and thresholds at a temperature of 260\,K. At each measured over-voltage the dark count rate at the 0.5\,PE threshold was extrapolated from this measurement. The 0.5\,PE threshold was measured for the different over-voltages by fitting the pulse height histogram from the gain calibration with a Gaussian function, as shown in Figure~\ref{fig:DCR_Sweeping}, right. The obtained dark count rate as a function of over-voltage at each temperature setting was fitted with a linear function, as shown in Figure~\ref{fig:DC_Results}, left. Figure~\ref{fig:DC_Results}, right, shows the dark count rate as a function of temperature for various over-voltages. The dark count rate decreases two orders of magnitude when going from room temperature to 230\,K and follows an exponential function, which is predicted by the Shockley-Read-Hall (SRH) carrier generation model~\cite{PhysRev.87.835}. The extracted points were fitted with an exponential function. We measured an activation energy of (0.52$\pm$0.20)\,eV, a value which is close to half of the band gap energy of silicon $E_g^\mathrm{Si} = 1.12$\,eV~\cite{doi:10.1063/1.1754731}. This implies that trap levels are present in the center of the silicon band gap, which acts as a dominant SRH carrier generation center. This effect was also observed in~\cite{Pagano:2012}. 

\begin{figure}[t!]
\centering
\includegraphics[height=5.3cm]{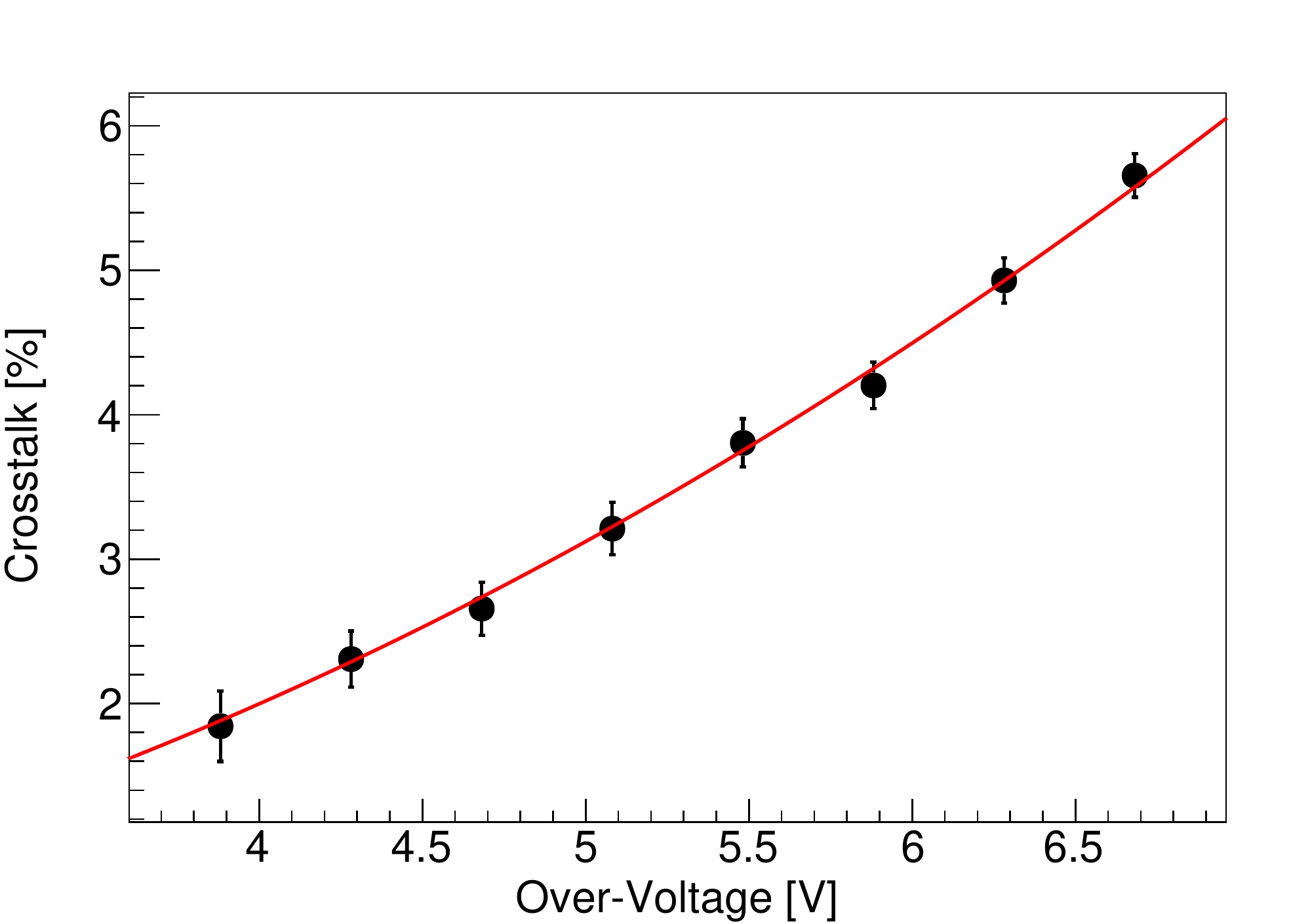}
\includegraphics[height=5.3cm]{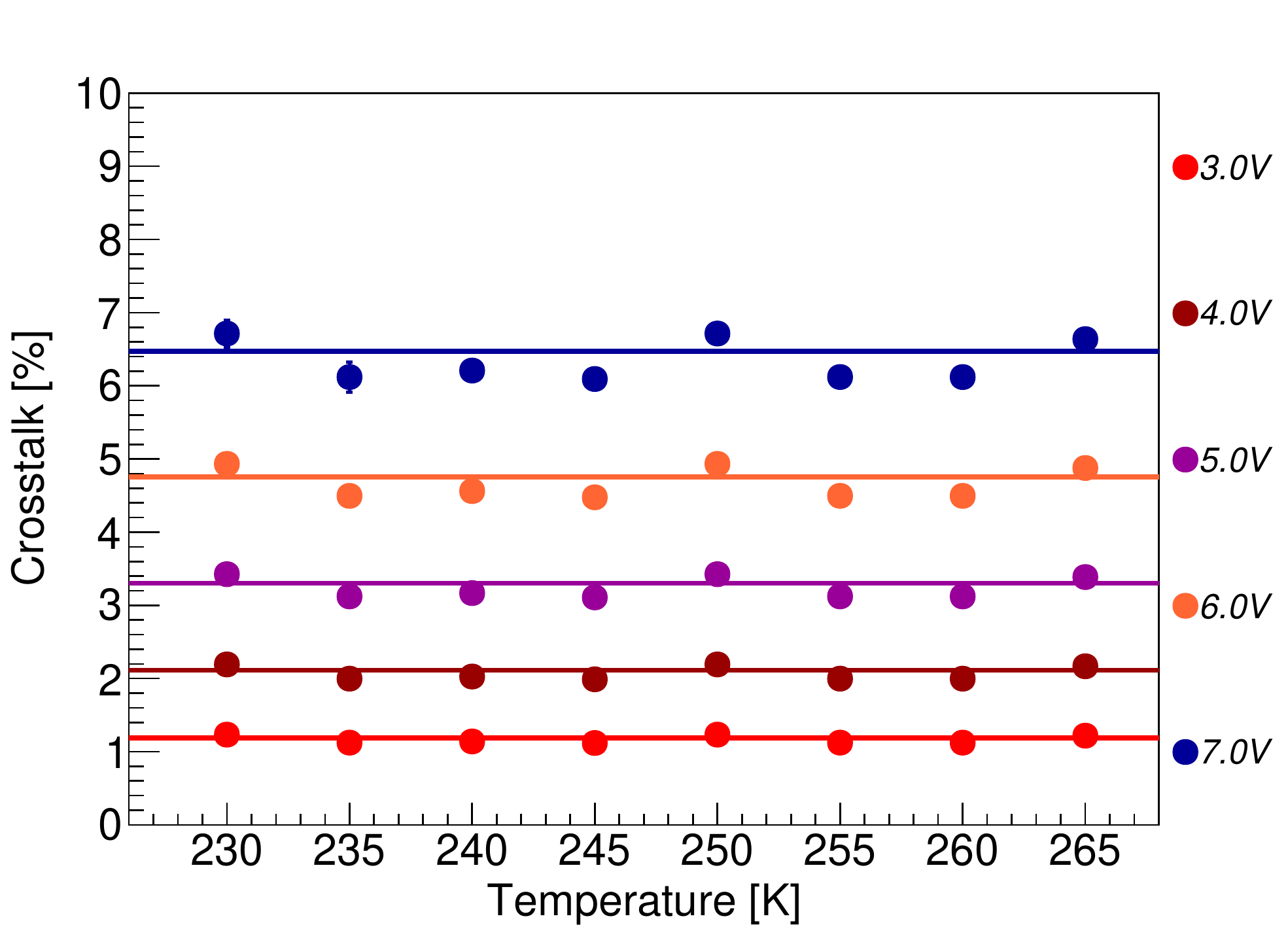}
\caption{ (Left): Crosstalk rate as a function of over-voltage at 260\,K. The points are fitted with a function in the form $a\times\Delta V^2$ (red line). (Right): The crosstalk rate as a function of temperature for different over-voltages, from 3 to 7\,V. The crosstalk rate remains constant with temperature.}
\label{fig:CT_Results}
\end{figure}

The crosstalk rate was measured by counting thermally generated single pulses at different bias voltages as a function of temperature at 0.5\,PE and 1.5\,PE thresholds. It is thus defined as the ratio $\frac{\mathrm{DCR}_{1.5\,\mathrm{PE}}}{\mathrm{DCR}_{0.5\,\mathrm{PE}}}$.  The $\mathrm{DCR}_{0.5\,\mathrm{PE}}$  and $\mathrm{DCR}_{1.5\,\mathrm{PE}}$ for different over-voltage were obtained by evaluating the dark count rate as a function of signal threshold at the corresponding pulse thresholds. The obtained crosstalk rate as a function of over-voltage at each temperature setting was fitted with a function  $a\times \Delta V^2$, where $a$ is a free parameter,  as shown in Figure \ref{fig:CT_Results}, left. Figure \ref{fig:CT_Results}, right, shows the crosstalk rates as a function of temperature for a set of  over-voltages. We observe that the crosstalk rate stays constant with temperature, as seen also in~\cite{Biroth:2015lxa}. This allows us to evaluate this parameter at room temperature. 

\subsection{Long-Term Stability in Cold Gaseous Nitrogen} 
\label{subsection:LArS_Longterm}

We measured the dark count rate and the gain of the SiPM at a mean temperature of 172\,K in gaseous nitrogen for 27 days. For this measurement, we controlled the temperature via the nitrogen flow through the copper coil, which resulted in less nitrogen consumption and a temperature rms of $\sim$\,0.14\,K. The SiPM was operated at an over-voltage of 5.6\,V for a good signal-to-noise ratio.

\begin{figure}[b!]
\includegraphics[height=5.3cm]{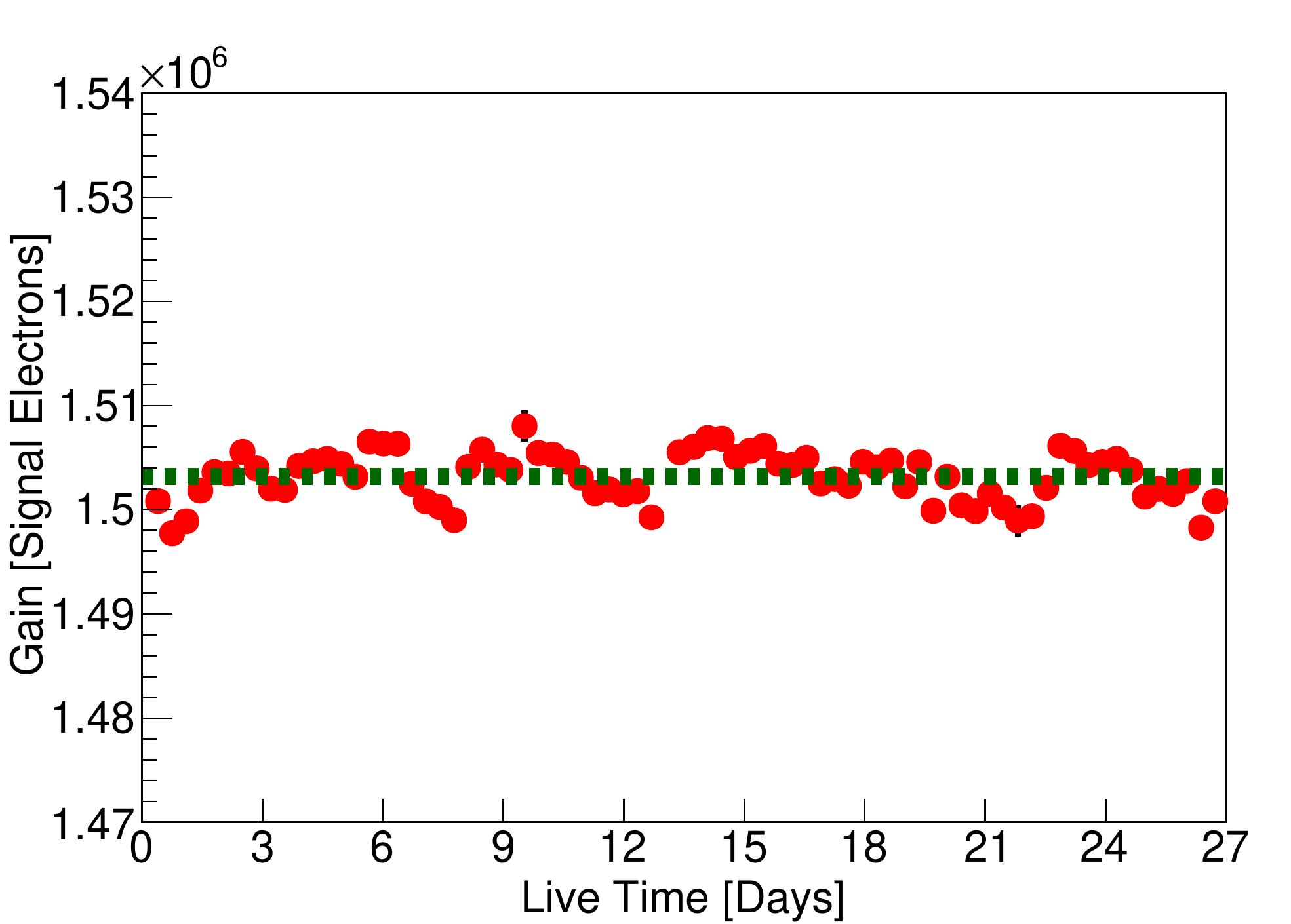}
\includegraphics[height=5.3cm]{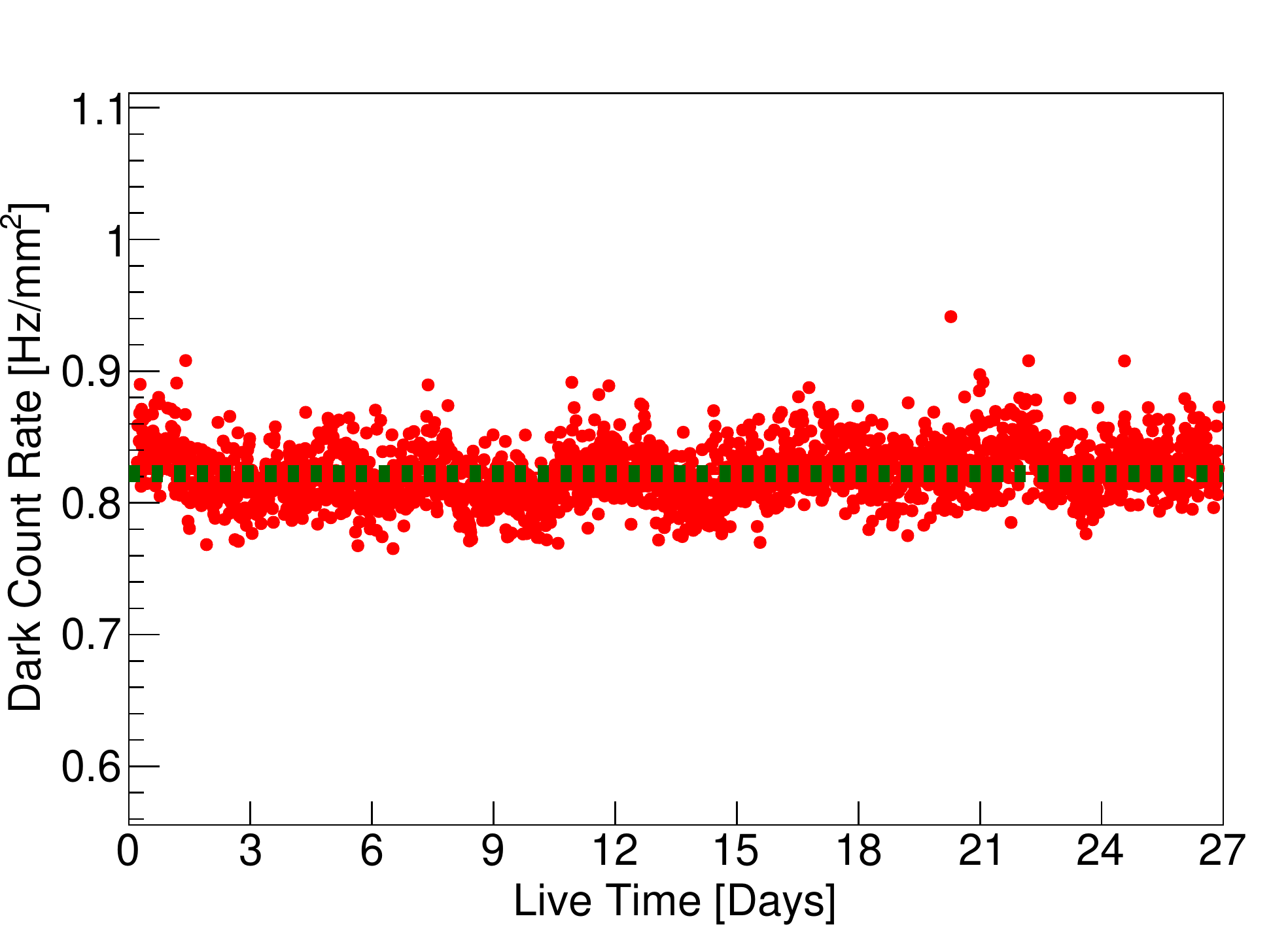}
\caption{(Left): Gain stability at a temperature of 172\,K as a function of time. The data (red points) was collected over a period of 27 days. The green dotted line indicates the mean gain value, which is $1.50\times10^6$ with an rms of $0.01\times10^6$.  (Right): Dark count rate (red data points) at 0.5\,PE signal height at a temperature of 172\,K. The green dotted line indicates the mean rate, which is 0.82\,Hz/mm$^2$ with an rms of 0.03\,Hz/mm$^2$.}
\label{fig:longterm_sipm}
\end{figure}

We continuously monitored the dark count rate, and we performed regularly a gain calibration with LED light. The gain evolution is shown in Figure~\ref{fig:longterm_sipm}, left. The measured mean value is $1.50\times10^6$ with an rms of $0.01\times10^6$. The gain fluctuations can be correlated with the temperature fluctuations due to the nitrogen dewar exchange and are below 1\%. The SiPM  shows a good long-term stability without any visible gain drift. The dark count rate at a 0.5\,PE threshold is shown in Figure \ref{fig:longterm_sipm}, right. It remains stable over the entire measurement period with a mean normalised rate per area of 0.82\,Hz/mm$^2$ and an rms of 0.03\,Hz/mm$^2$. At a similar temperature, PMTs show an average dark count rate  of 0.01\,Hz/mm$^2$~\cite{Barrow:2016doe}. Therefore, further developments by the manufacturer are necessary in order to achieve comparable dark count rates.

\section{SiPM Operation in Liquid Xenon}
\label{sec:liquid_xenon_operation}

\begin{figure}[b!]
	\includegraphics[height=5.3cm]{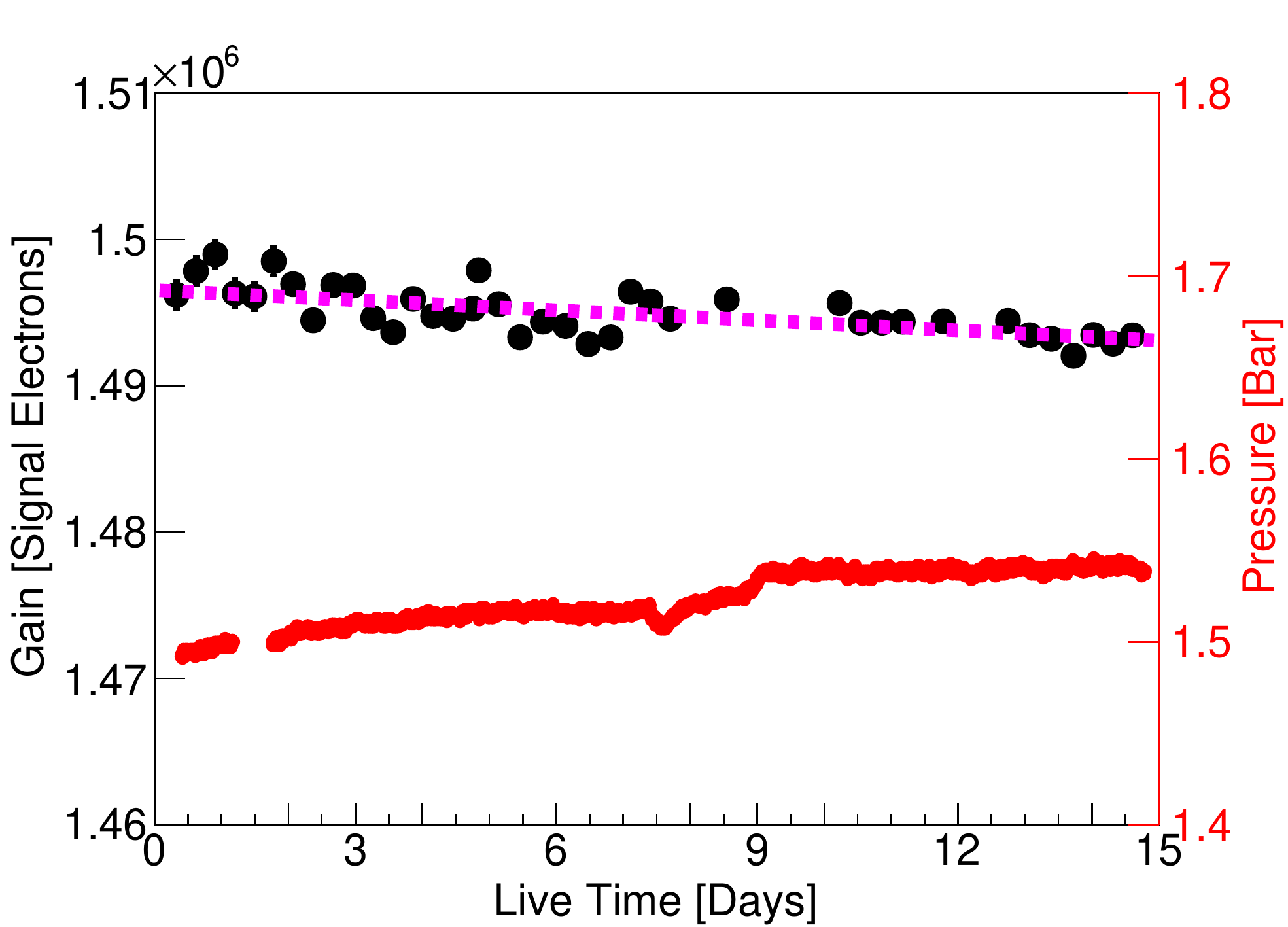}
	\includegraphics[height=5.3cm]{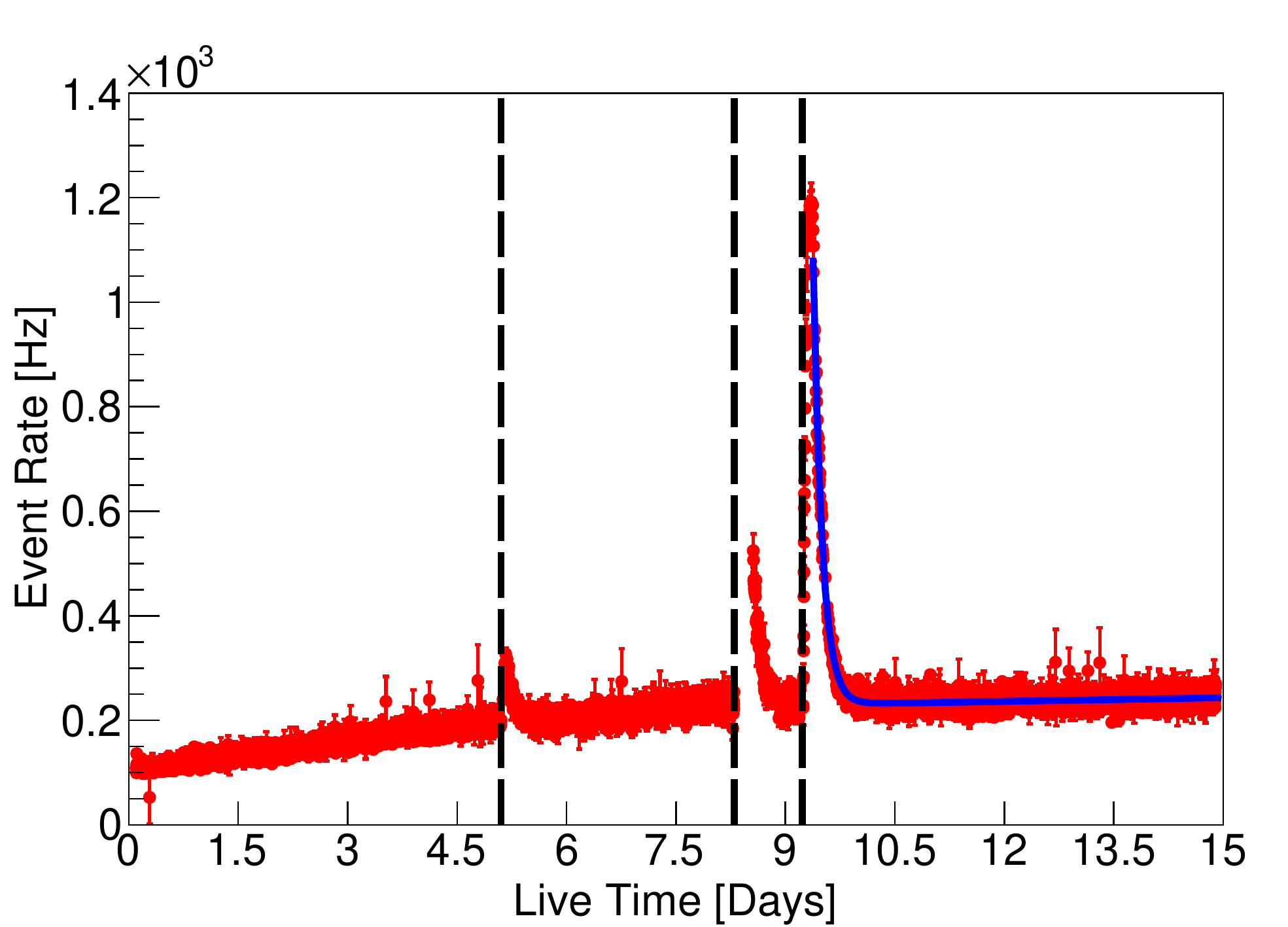}
	\caption{(Left): The gain of the SiPM in liquid xenon (black points) and pressure inside the setup (red points) as a function of time for 15 days. The gain, fitted with a linear function (magenta dashed line) shows a slight decrease, which is correlated to the temperature and pressure increase in the detector with time. (Right): The event rate measured by the SiPM at 0.5\,PE signal height. It increases with time, due to the continuous purification of the xenon gas and hence increase in light yield. We injected three times $^{83\mathrm{m}}$Kr into the setup (black dashed lines), visible by the three peaks above the smooth distribution. By fitting the event rate after the injection with an exponential combined with a linear function (blue curve), to model the background events, we obtain the expected $^{83\mathrm{m}}$K-decay time of ($1.86\pm0.03$)\,h.}
	\label{fig:longterm_lxe}
\end{figure}In the first part of this section we demonstrate the long-term performance of the SiPM in liquid xenon. 
	To verify the sensitivity to xenon scintillation light with a wavelength of 178\,nm, we injected the $^{83\mathrm{m}}$Kr source and measured its decay time. In the second part we report the results of our long-term calibration with the $^{241}$Am source, where we monitored the light yield of the xenon chamber.
The measurements presented in this section were performed in a single-phase liquid xenon detector  (MarmotX) described in~\cite{Barrow:2016doe}. The setup was optimised to evaluate the XENON1T and XENONnT PMTs in liquid xenon. The SiPM was positioned near the center of the inner detector facing a small PTFE chamber containing a $^{241}$Am calibration source (described in \cite{Baudis:2015saa}) with an $\alpha$-particle energy of 5.49\,MeV and a half-life of 432 years. We kept the same readout chain as for the previous measurements. The setup was filled with xenon and the SiPM was completely submerged in the liquid at a pressure of 1.5\,bar. The xenon was constantly purified by circulating it through a hot metal getter (at a flow rate of 4\,SLPM) and the gas system was equipped with a mixing chamber that allowed us to introduce the metastable $^{83\mathrm{m}}$Kr calibration source  (described in \cite{Manalaysay:2009yq}) into the liquid.

We measured the event rate and the gain of the SiPM in liquid xenon for 15 days. The temperature of the setup increased linearly during this time by $\sim$1\,K, resulting in a pressure increase of $\sim$0.05\,bar. For this measurement, we operated the flash ADC in self-trigger mode with a constant threshold of 0.5\,PE. We regularly acquired the single photoelectron spectrum and extracted the gain using the same procedure as described in section \ref{subsection:spe}. The long-term stability of the gain and the pressure inside the setup is shown in Figure \ref{fig:longterm_lxe}, left. We obtained a mean gain value of 1.5$\times10^6$ with an rms of $0.02\times10^6$. The gain shows a slight decrease, which is correlated to the increase of the breakdown voltage, due to the above mentioned temperature and pressure increase in the setup.

\begin{figure}[b!]
	\includegraphics[height=5.4cm]{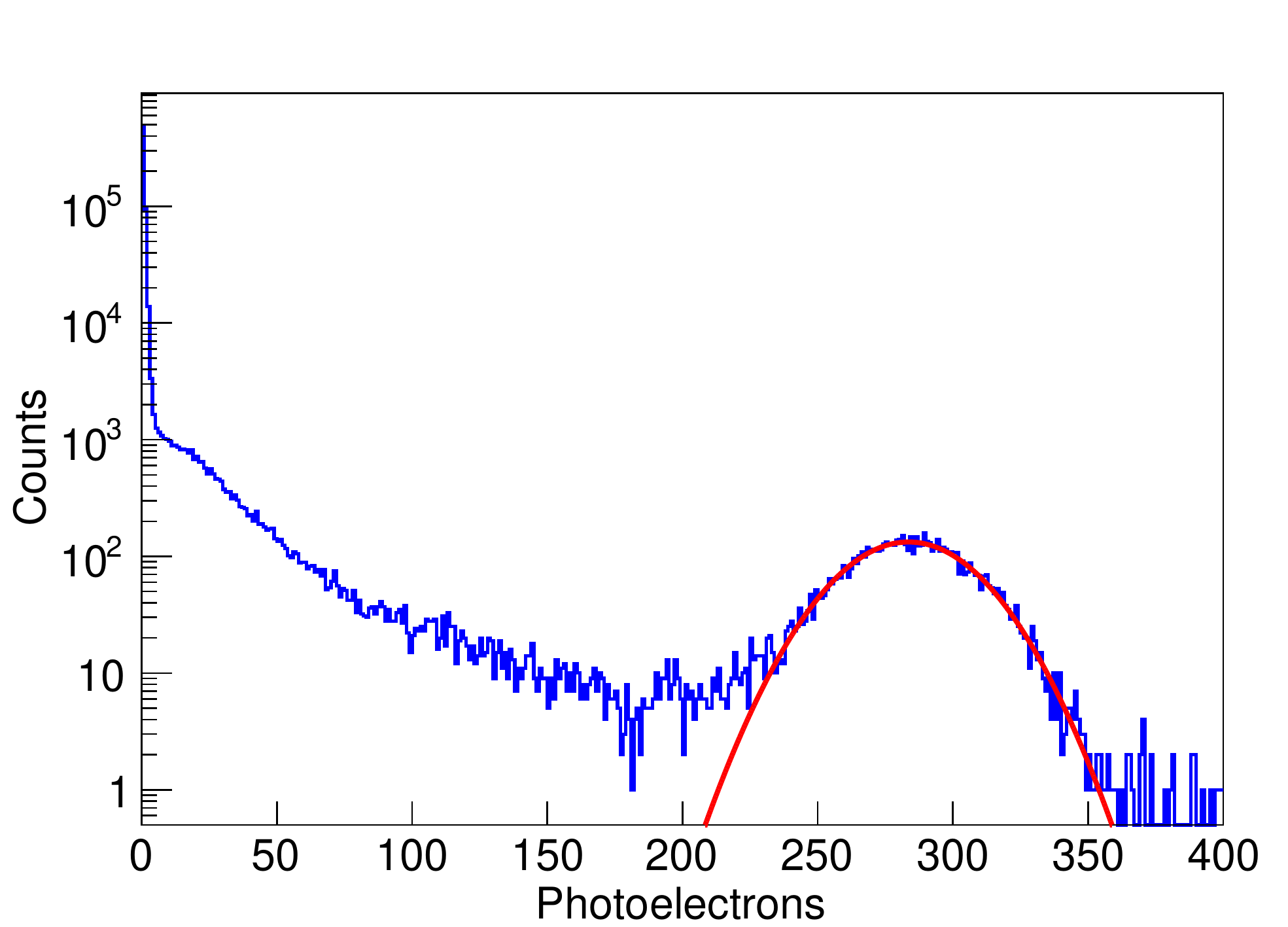}
	\includegraphics[height=5.4cm]{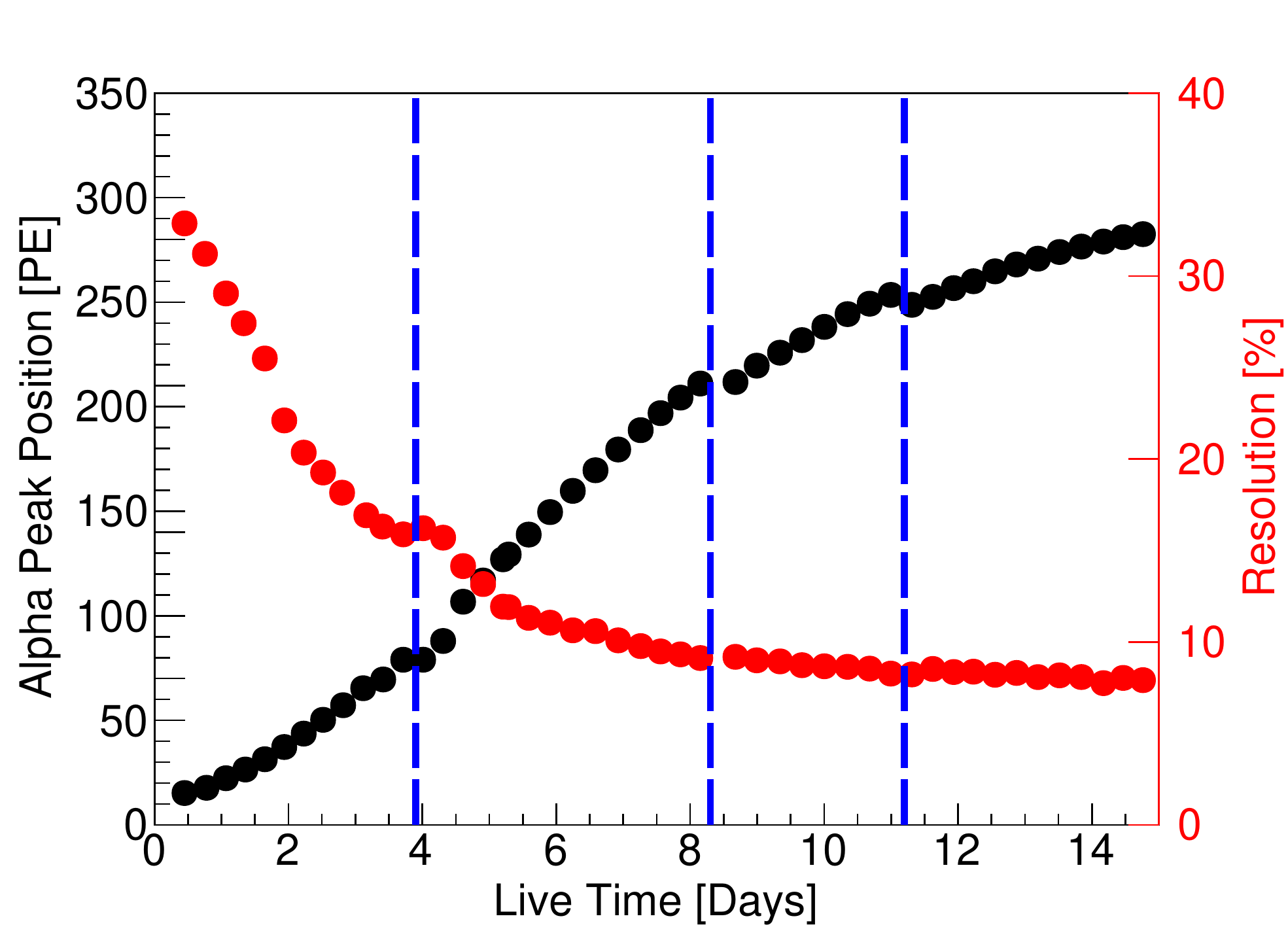}
	\caption{(Left): A measured SiPM spectrum (blue histogram) with the $^{241}$Am source in liquid xenon. At high energies the $\alpha$-peak is clearly visible and fitted with a Gaussian function (red curve) to obtain the mean $\alpha$-peak position. (Right): The mean $\alpha$-peak position and resolution over time for 15 days. The measured position increases due to the continuous purification of the xenon gas, and hence this leads to an increase in the light yield. The observed periods with a decrease in the alpha peak position can be correlated with times when we stopped the xenon gas recirculation (blue dashed lines).}
	\label{fig:alpha_peak}
\end{figure}

The event rate was measured at a threshold of 0.5\,PE, and its behaviour as a function of time is shown in Figure \ref{fig:longterm_lxe}, right. This rate consists of thermally generated pulses in the SiPM and pulses due to the detected xenon scintillation light. The observed rate increases over the measurement period due to the continuous purification of the xenon and thus an increase in the light yield.  We injected the $^{83\mathrm{m}} $Kr source at three different times into the gas system and hence into the liquid, visible in the three peaks above the smooth distribution. The event rate differs among these peaks, due to different $^{83\mathrm{m}} $Kr injection durations. In the last $^{83\mathrm{m}}$Kr injection we waited until the rate stabilised at a constant level,  to achieve a homogeneous event distribution inside the detector.  We extracted the $^{83\mathrm{m}}$Kr decay time by fitting the event rate after the third $^{83\mathrm{m}} $Kr injection with an exponential combined with a linear function, which models the increase of the background event rate due to the xenon purification. The fit yields a decay time of ($1.86\pm0.03$)\,h. This value is in agreement with previous measurements~\cite{Kastens:2009pa} and consistent with the published value of $1.83$\,h~\cite{Browne:110620}.

We measured the light yield evolution in the liquid xenon setup for 15 days. For this purpose, we continuously monitored the xenon scintillation light produced by $\alpha$-decays of a $^{241}$Am \mbox{$\alpha$ source}. We regularly acquired a charge spectrum with the ADC, as shown in Figure \ref{fig:alpha_peak}, left, for one particular measurement. We converted the number of signal electrons to photoelectrons by using the measured gain, shown in Figure  \ref{fig:longterm_lxe}, left. The $\alpha$-peaks were fitted with Gaussian functions to obtain their mean position. Figure \ref{fig:alpha_peak}, right, shows the evolution of the  $\alpha$-peak position in time. The  $\alpha$-peak position is continuously shifting to higher values due to xenon gas purification and thus the increase in the overall light yield. We stopped  the recirculation through the getter three times, due to the $^{83\mathrm{m}}$Kr  injections, which is visible as a decrease in the $\alpha$-peak position. After 15 days of measurement, we reached a light yield of (45.29 $\pm$0.16)\,PE/MeV. The energy resolution computed on the  $\alpha$-peak improved from 32\% at the start to 8\% at the end of the measurement.


\section{Radiopurity Measurements of the SiPM Components}
\label{sec:Radiopurity}


Rare-event search experiments demand an ultra-low background in order to achieve high sensitivity, calling for the use of detector materials with very low radioactivity levels. This requirement is particularly applicable to the photosensors, which are placed close to the active target for detecting interactions with high efficiency. The SiPM raw materials from Hamamatsu were investigated for traces of radioactive isotopes with high-purity germanium (HPGe) spectrometers and inductively coupled plasma mass spectrometry (ICP-MS). In Table \ref{table:material} we list the investigated components, the content of each material by mass per 12$\times$12\,mm$^2$ SiPM, and the screening exposure of each sample for the HPGe measurements.

 \begin{table}[h!]
 \begin{center}
\begin{tabular}{lcccc}
{Sample type}  & Mass/SiPM [g]&{ Measured mass [g]} &{ Observation time [days]} \\
\hline
Silicon chips  & 0.10 &  104.06 & 64.0  \\
Bonding resin, type C & 0.015&9.6  & 34.3  \\
Bonding resin, type D & 0.004 &12.6  & 44.0  \\
Quartz window  & 0.30  & 14.2 &45.2   \\
Quartz packaging  & 0.31 &  19.1 &32.7  \\
\end{tabular}
\end{center}
\caption{List of measured raw material samples including the mass of each component per 12$\times$12\,mm$^2$ SiPM, the measured mass and the counting time. The total mass of the SiPM is 0.73\,g.}
\label{table:material}
\end{table}

The HPGe spectrometers within the GeMPI~\cite{PMID:10879860} and  Gator~\cite{Baudis:2011am} facilities, located underground at the Laboratori Nazionali del Gran Sasso (LNGS), were used for the sample measurements. These low-background detectors have a high energy resolution and provide a non-destructive testing method that is sensitive to almost every gamma-ray emitting isotope. The presence of isotopes in the primordial uranium and thorium chains as well as the gamma-emitter $^{40}$K are determined through spectroscopic analysis, yielding sensitivities of a few mBq/kg for most of the main gamma lines, depending on the sample exposure. For enhanced sensitivity to the $^{238}$U and $^{232}$Th content of the silicon chips, an ICP-MS measurement was performed by the chemistry laboratory at LNGS. For these primordial isotopes, sensitivities on the order of 1-100\,$\mu$Bq/kg can typically be achieved through ICP-MS, with a measurement uncertainty between 20-30$\%$. The methodology and sample preparation used for these measurements are identical to those described in~\cite{XENON1T_radioassay}.

The results of these measurements are listed in Table \ref{table:results} for each sample, weighted by component mass per SiPM. In addition, the activity of each weighted component is normalised by the SiPM sensor area. The number of the total SiPM activity is a conservative estimate, as for most cases only upper limits were obtained for the raw materials. We compare these first SiPM results to the average activities of the Hamamatsu R11410-21 PMTs, that was developed for low radioactivity and currently used in the XENON1T dark matter experiment~\cite{XENON1T_PMTradioactivity,XENON1T_radioassay}.   

\begin{table}[h!]
\begin{center}
\begin{tabular}{l|cccccc}
Sample type &{ $^{238}$U} & {$^{226}$Ra}  &  {$^{228}$Ra ($^{232}$Th$^{*}$)}  &  { $^{228}$Th} &  { $^{40}$K } \\
  \hline
Silicon chips    &  $<0.002^{*}$ & $<0.0003$ & $<0.00007^{*}$ & $0.0004(1)$ & $ <0.0014 $  \\
Bonding resin, type C   &  $ <0.299$ & $0.0043(9)$ & $  <0.003 $ & $ <0.003 $ & $ 0.02(1)$  \\
Bonding resin, type D   &  $  <0.588$ & $<0.0027$ & $<0.006$ & $0.003(1)$ & $<0.00004$ \\
Quartz window    &  $<0.013$ & $0.00009(3)$ & $<0.00001$ & $ 0.0001(3)$ & $0.004(2)$ \\
Quartz packaging    &  $<0.006$ & $0.00011(1)$ & $0.00011(2)$ & $0.0001(2)$ & $<0.0001$ \\
\hline
Total SiPM  &   $<0.908$ & $<0.0075$ & $<0.0092$ & $<0.0066$ & $<0.026$\\
Total  R11410 PMT    &  $<0.4$  &  $0.016(3)$ &  $0.016(4)$ & $0.012(3)$  & $0.37(6)$  \\
\end{tabular}
\end{center}
\caption{SiPM radioassay results from gamma-ray spectroscopy for the uranium and thorium decay chains, as well as for the $^{40}$K single isotopes in mBq/cm$^2$. Results are given at 90$\%$ confidence level for upper limits of the silicon and quartz samples. Upper limits for the resin samples are given at 68$\%$ confidence level. The ICP-MS results (indicated by $^{*}$) were obtained with a 90$\%$ recovery efficiency. Detections for both SiPM samples and PMTs~\cite{XENON1T_radioassay} are given with uncertainties of $\pm1\sigma$.}
\label{table:results}
\end{table}

From the results of Table \ref{table:results} we see that the SiPMs are promising for use in low-background experiments. The achieved upper limits for the SiPM materials are already comparable to or an order of magnitude lower than the PMT activities; the latter results were achieved only after several optimisations of the R11410-21 bulk material for radiopurity. The next steps in characterising SiPM for use in low-background experiments is to improve the sensitivity of the results, particularly for the upper part of the uranium decay chain, by increasing the sample sizes for HPGe measurements and by obtaining additional ICP-MS measurements. Further optimisation of some of the raw materials for lower radioactivity, such as the resins, is necessary.

\section{Conclusions and Outlook}
\label{sec:conclusions}

We operated a 6$\times$6\,mm$^2$ SiPM unit from Hamamatsu in gaseous nitrogen at temperatures down to 110\,K, as well as in liquid xenon at a  temperature of 186\,K for 15 days. We list the main SiPM performance parameters at 172\,K in Table~\ref{table:SiPM_Parameters}. The SiPM showed a stable operation at gain of $1.5\times10^6$, which is comparable to  state-of-the-art R11410-21 PMTs~\cite{Barrow:2016doe}.  We showed that the dark count rate decreases by several order of magnitudes when going from room temperature to 172\,K, reaching 0.8\,Hz/mm$^2$ at an over-voltage of 5.6\,V. These rates are still a factor of 80 higher than those achieved in cryogenic PMTs operated at the same temperature, requiring further developments.  We measured the crosstalk rate and showed that it is independent of temperature. We also demonstrated a stable long-term operation of the device both in cold nitrogen gas, and in liquid xenon. No significant changes in gain were observed and the dark count rate was stable in time as well. We tested the SiPM with xenon scintillation light using an internal $^{83\mathrm{m}}$Kr source mixed with the xenon, where the measured half-life agrees with other measurements, and with $\alpha$ particles from a $^{241}$Am source. Finally, we performed radio-purity measurements of SiPM material components, which showed that SiPMs are promising candidates for photosensor in liquid xenon detectors.\begin{table}[h!]
	\centering
	\begin{tabular}{ccccc}
		Over-voltage [V] & Gain & SPE Resolution [\%]& Dark count rate [Hz/mm$^2$] & Crosstalk [\%]  \\
		\hline
		5.6 &  1.5$\times10^6$ & 9 &0.8 & 4.1 \\
	\end{tabular}
	\caption[Measurement results of the VUV sensitive SiPM]{ Summary of the main 6$\times$6\,mm$^2$ SiPM performance parameters at 172\,K and a specific over-voltage.}
	\label{table:SiPM_Parameters}
\end{table}

For an upgrade of the Xurich~\cite{Baudis:2017xov} xenon time projection chamber, we are replacing the top and bottom PMTs with SiPM arrays to test their behaviour in a TPC environment. This is the next step towards probing their suitability for larger liquid xenon experiments such as DARWIN~\cite{Aalbers:2016jon}. In addition, we plan to build a larger TPC with 4$\pi$ SiPM readout, to explore the possibility of replacing wall reflectors with active photosensor units.

\section*{Acknowledgements}

This work was supported by the Swiss National Science Foundation under Grant No. 200020-175863, by the European Unions Horizon 2020 research and innovation programme under the Marie Sklodowska-Curie grant agreements No. 690575 and No. 674896, and by the European Research Council (ERC) under the European Union's Horizon 2020 research and innovation programme, grant agreement No. 742789 (Xenoscope). We thank Matthias Laubenstein for the $\gamma$-spectrometry measurements with the GeMPI detectors, and Stefano Nisi for the ICP-MS measurements at LNGS. We also thank Andreas James for significant contributions to the design and construction of the evaluation setups and its sub-systems, and David Wolf for his help in the design and production of the printed readout circuit boards.


\end{document}